\documentclass{aa}

\usepackage{graphicx}
\usepackage{txfonts}
\usepackage[breaklinks,colorlinks,citecolor=blue]{hyperref}

\usepackage{natbib}
\bibpunct{(}{)}{;}{a}{}{,}

\newcommand\be{\begin{equation}}
\newcommand\ee{\end{equation}}
\newcommand\bea{\begin{eqnarray}}
\newcommand\eea{\end{eqnarray}}
\def\ba#1\ea{\begin{align}#1\end{align}}

\newcommand\dd{\mathrm{d}}

\newcommand\mr{\mathrm}
\newcommand\lbra{\left\langle}
\newcommand\rbra{\right\rangle}
\newcommand\Cov{\mr{Cov}}

\newcommand\Clgal{\ensuremath{C_\ell^\mathrm{gal}}}
\newcommand\nbargal{\ensuremath{\overline{n}_\mathrm{gal}}}

\newcommand{\threeJz}[3]{\begin{pmatrix} #1 & #2 & #3 \\ 0 & 0 & 0 \end{pmatrix}}

\newcommand{\Euclid}{\textit{Euclid}}

\newcommand{\uuid}{2cdfb740-27c9-4e91-bec4-ef0f6b4835ad}

\begin{document}

\title{The impact of braiding covariance and in-survey covariance on next-generation galaxy surveys}
\titlerunning{Braiding and in-survey covariance}

\author{Fabien Lacasa\thanks{fabien.lacasa@universite-paris-saclay.fr}\inst{\ref{inst1}} }
\institute{
Université Paris-Saclay, CNRS, Institut d'astrophysique spatiale, 91405, Orsay, France.\label{inst1}
}

\date{\today}

\abstract
{
As galaxy surveys improve their precision thanks to lower levels of noise and the push toward small, non-linear scales, the need for accurate covariances beyond the classical Gaussian formula becomes more acute. Here I investigate the analytical implementation and impact of non-Gaussian covariance terms that I had previously derived for the galaxy angular power spectrum. 
Braiding covariance is such an interesting class of such terms and it gets contributions both from in-survey and super-survey modes, the latter proving difficult to calibrate through simulations. I present an approximation for braiding covariance which speeds up the process of numerical computation. I show that including braiding covariance is a necessary condition for including other non-Gaussian terms, namely the in-survey 2-, 3-, and 4-halo covariance. Indeed these terms yield incorrect covariance matrices with negative eigenvalues if considered on their own.
I then move to quantify the impact on parameter constraints, with forecasts for a survey with Euclid-like galaxy density and angular scales.
Compared with the Gaussian case, braiding and in-survey covariances significantly increase the error bars on cosmological parameters, in particular by 50\% for the dark energy equation of state $w$. The error bars on the halo occupation distribution (HOD) parameters are also affected between 12\% and 39\%. Accounting for super-sample covariance (SSC) also increases parameter errors, by 90\% for $w$ and between 7\% and 64\% for HOD. In total, non-Gaussianity increases the error bar on $w$ by 120\% (between 15\% and 80\% for other cosmological parameters) and the error bars on HOD parameters between 17\% and 85\%.
Accounting for the 1-halo trispectrum term on top of SSC, as has been done in some current analyses, is not sufficient for capturing the full non-Gaussian impact: braiding and the rest of in-survey covariance have to be accounted for. Finally, I discuss why the inclusion of non-Gaussianity generally eases up parameter degeneracies, making cosmological constraints more robust for astrophysical uncertainties. I release publicly the data and a Python notebook reproducing the results and plots of the article at \url{https://github.com/fabienlacasa/BraidingArticle}.
}


\maketitle


\section{Introduction}\label{Sect:intro}

With the increase of galaxy density in current and forthcoming cosmic surveys, our statistical analysis of the large-scale structure of the Universe needs to be pushed towards new degrees  of precision. Accurate covariance matrices are an important part of this effort. Indeed, using an incorrect covariance basically amounts to analysing a biased data set \citep[e.g.][]{Sellentin2019}. This effect has, indeed, been seen in current weak lensing surveys, along with changes in the covariance shifting cosmological constraints on $S_8 = \sigma_8 (\Omega_m/0.3)^{0.5}$ \citep{Hildebrandt2017,Troxel2018}, which is of particular importance in the current context of possible tensions between low- and high-redshift measurements of $\sigma_8$.

In the past, covariance matrices for the large scale structure were often estimated using jackknife or bootstrap techniques, however, this has been shown to be biased at a level inadequate for cosmological analyses \citep{Norberg2009,Lacasa2017}. 
Other analyses have used matter covariances coming from ensemble of simulations \citep{Harnois-Deraps2013}, which correctly capture  in-survey covariance, projected in 2D using the flat sky approximation \citep{Sato2013}. These matter covariances cannot be directly applied, however, to galaxy clustering, as \cite{Abramo2015} highlighted in stating that the amount of non-linearity is notably dependent on the galaxy selection; the scaling with bias of a 1-halo polyspectrum  differs greatly from the perturbative polyspectrum.
Most current analyses use covariances that either come from analytical computations or from dedicated simulations. In particular, analytical modeling using the halo model has risen to a state of the art level for several current galaxy surveys \citep{Krause2017,Hildebrandt2017,Krause2017b}. This is the approach followed in this article, applying it to a galaxy clustering analysis using the angular power spectrum. I emphasise that the analysis and conclusions can be transferred to a real-space analysis using the two-point correlation function since there is a linear mapping between real space and harmonic space, where computations are much simpler to carry out.

The point of halo model covariances is to move beyond the vanilla analytical Gaussian formula. In the case of angular auto-spectra for galaxy clustering in disjoint redshift bins labeled $i_z,j_z$, this formula gives
\be
\mathcal{C}_{\ell,\ell'}^G = \frac{2 \ C_\ell^\mr{gal}(i_z)^2}{2\ell+1} \ \delta_{\ell,\ell'} \ \delta_{i_z,j_z} ,
\ee
where throughout the article I use the short notation
\be
\mathcal{C}_{\ell,\ell'} \equiv \Cov\left(C_\ell^\mr{gal}(i_z),C_{\ell'}^\mr{gal}(j_z)\right) .
\ee
Using the halo model for non-Gaussian covariance terms allows not only for an adequate reproduction of super-sample covariance \citep[SSC,][]{Takada2013} for power spectra, but it further allows for the inclusion of one-point statistics, such as cluster counts \citep{Lacasa2016}, and 3-point statistics, such as the weak lensing bispectrum \citep{Rizzato2019}, as well.

Here I build upon \cite{Lacasa2018b} and  its exhaustive analytical derivation of the covariance of the galaxy angular power spectrum with all non-Gaussian terms. Specifically, I implement the terms derived and argued there as potentially being of importance and I gauge their impact on the information content of galaxy clustering. To this end, I also use the halo model at tree-level for the prediction of the observable $\Clgal$ and of the covariance $\mathcal{C}_{\ell,\ell'}$.

In detail, I first carried out an analytical study of non-Gaussian covariance terms of the galaxy angular power spectrum (Sect.~\ref{Sect:analytical-cov}), recalling the analytical expressions from \cite{Lacasa2018b} (Sect.~\ref{Sect:NGterms}). Then I\ presented an approximation for braiding covariance, making it numerically tractable (Sect.~\ref{Sect:braiding-approx}). I then presented numerical results that first demonstrate the importance of braiding covariance (Sect.~\ref{Sect:cov-setup}) and then show analytically that accounting for braiding covariance is necessary for the inclusion of other in-survey covariance terms, such as 2h1+3, which have important off-diagonal contributions (Sect.~\ref{Sect:braiding-necessity}). I present a signal-to-noise analysis that shows that braiding and in-survey covariance have a substantial impact compared to a Gaussian covariance, although the impact is milder once super-sample covariance is also included (Sect.~\ref{Sect:SNR}). Afterwards, I move to a Fisher analysis to show the impact of non-Gaussianity on parameter constraints, both for cosmology (Sect.~\ref{Sect:impact-cosmo}) and for halo occupation distribution (HOD, Sect.~\ref{Sect:impact-HOD}). Finally, I discuss the results in Sect.~\ref{Sect:discu} and, in particular, I\ consider how parameter degeneracies are generally eased up by the inclusion of non-Gaussianity.

At \url{https://github.com/fabienlacasa/BraidingArticle} the data and a Python notebook that allows to reproduce all plots and results of the article are available, along with a bit more information.

\section{Analytical covariance}\label{Sect:analytical-cov}

In this section, I first set out the equations for the non-Gaussian covariance terms, then I present a numerical approximation for the specific case of braiding covariance.
For this purpose, a few definitions and notations are needed.

First, the (unobservable) angular power spectrum of matter between two redshifts $(z_a, z_b)$ is
\ba
C_\ell^m(z_a,z_b)= \frac{2}{\pi} \int k^2\,\dd k \ P(k|z_a, z_b) \ j_\ell(k r_a) \ j_\ell(k r_b) .
\ea
Second, halo model equations can be greatly simplified by introducing the integral
\ba
\nonumber I_\mu^\beta(k_1,\cdots,k_\mu|z) = \int \dd M \ & \frac{\dd n_h}{\dd M} \ \lbra N_\mr{gal}^{(\mu)}\rbra \ b_\beta(M,z) \\ 
& \times u(k_1|M,z) \cdots u(k_\mu|M,z)  
\ea
where $\frac{\dd n_h}{\dd M}$ is the halo mass function, $u(k|M,z)$ is the normalised halo profile, $b_\beta(M,z)$ is the halo bias or order $\beta$\footnote{The terms considered here only involve local bias up to the third order, so $\beta=0,1,2,3$ with $b_0=1$. More generally, we could have non-local bias such as coming from the tidal field at second order: $b_{s^2}$.}, and $\lbra N_\mr{gal}^{(n)}\rbra \equiv \lbra N_\mr{gal} (N_\mr{gal}-1)\cdots(N_\mr{gal}-(n-1))\rbra $ is the number of $n$-tuples of galaxies, implicitly depending on halo mass. \\
Finally, further simplifications can be achieved by grouping integrals together:
\ba
I_\mu^{\Sigma_2} = \frac{17}{21} \ I_\mu^{1} + \frac{1}{2!} \ I_\mu^{2}
\ea
is the sum of second order contributions from perturbation theory and local bias, and
\ba
I_\mu^{\Sigma_3} \equiv \frac{1023}{1701} \ I_\mu^{1} + \frac{1}{3!} \ I_\mu^{3}
\ea
is the sum of third order contributions \citep{Lacasa2018b}.

\subsection{Non-Gaussian terms}\label{Sect:NGterms}

I recapitulate the equations for all the non-Gaussian covariance terms so that this article may be self-contained. The equations all stem from \cite{Lacasa2018b}, with the slight modification that they are for the power spectrum of the usual galaxy density contrast, that is, $\Clgal \equiv C_\ell(\delta_\mr{gal})$, instead of the absolute power spectrum $C_\ell(n_\mr{gal})$ used in \cite{Lacasa2018b}. This is done to maintain maximal familiarity for most readers. In practice, this just changes an overall factor for power spectra and covariances and it does not change parameter constraints that are presented later in Sect.~\ref{Sect:impact-constraints} nor any of the conclusion on the importance of the various terms.

The first non-Gaussian covariance term is by far the most studied \citep[e.g.][]{Takada2013,Li2014,Li2014b,Lacasa2016,Lacasa2017,Lacasa2018,Li2018,Akitsu2017,Barreira2018} and the one whose impact is already well recognised even for some current surveys \cite[e.g.][]{Hildebrandt2017}: super-sample covariance (hereafter, SSC). It takes the form of
\ba\label{Eq:SSC}
\nonumber \mathcal{C}_{\ell,\ell'}^\mr{SSC} = \int \dd V_a \, \dd V_b \ \Psi_\ell^\mr{sqz}(z_a) \ \Psi_{\ell'}^\mr{sqz}(z_b) \ \sigma^2(z_a,z_b) \ \Bigg/ \ N_\mr{gal}(i_z)^2 \, N_\mr{gal}(j_z)^2 ,
\ea
where $z_a \in i_z$, $z_b \in j_z$, $\dd V_a = r^2(z_a) \frac{\dd r}{\dd z}(z_a)$,

and\ba
\sigma^2(z_a,z_b) = \frac{C_0^m(z_a,z_b)}{4\pi}
\ea
is the SSC kernel, and with the angle-independent trispectrum terms from the halo model, \cite{Lacasa2018b} find
\ba
\Psi_\ell^\mr{sqz}(z) = 4 \ I_1^{\Sigma_2}(k_{\ell}|z) \ I_1^1(k_{\ell}|z) \ P(k_{\ell}|z) + I_2^1(k_{\ell},k_{\ell}|z) ,
\ea
which can be related to the more usual power spectrum response $\frac{\partial P_\mr{gal}}{\partial \delta_b}$ via $\Psi_\ell^\mr{sqz}(z) = \frac{\partial P_\mr{gal}(k_\ell)}{\partial \delta_b} \times I_1^0(z)^2$.\\
Although a fast approximation to SSC was recently presented by \cite{Lacasa2019}, I prefer  to maintain an exact computation here. I checked that the quick approximation gives results within 5\% to that of the full computation  of Eq.~\ref{Eq:SSC} for all numerical results presented throughout the article.

Next we have non-Gaussian terms, coming from the diagonal-independent part of the trispectrum. The first and simplest is the 1-halo term where all galaxies of the 4-point function reside in the same halo,
\ba\label{Eq:1halo}
\mathcal{C}_{\ell,\ell'}^\mr{1h} = \frac{\delta_{i_z,j_z}}{4\pi} \int \dd V \ I_4^0(k_{\ell},k_{\ell},k_{\ell'},k_{\ell'}|z) \ \Bigg/ \ N_\mr{gal}(i_z)^4 .
\ea
Then come higher halo terms which should not be included independently, as I show in Sect.~\ref{Sect:braiding-necessity}. We have the 2-halo 1+3 term, where one galaxy sits in a halo and the three others sit in another halo,
\ba\label{Eq:2halo1+3}
\nonumber \mathcal{C}_{\ell,\ell'}^\mr{2h1+3} = \frac{2\delta_{i_z,j_z}}{4\pi} & \int \dd V \ I_1^1(k_\ell|z) \ I_3^1(k_\ell,k_{\ell'},k_{\ell'}|z) \ P(k_\ell|z) \ \Bigg/ \ N_\mr{gal}(i_z)^4 \\
& + (\ell \leftrightarrow \ell') \ ,
\ea
the 3-halo base term,
\ba\label{Eq:3halobase}
\nonumber \mathcal{C}_{\ell,\ell'}^\mr{3h-base0} = \frac{\delta_{i_z,j_z}}{4\pi} & \int \dd V \ 2 \;  \left(I_1^{1}(k_\ell|z) \ P(k_\ell|z)\right)^2 I_2^{\Sigma_2}(k_{\ell'},k_{\ell'}|z) \Bigg/ N_\mr{gal}(i_z)^4 \\
\nonumber & + \quad (\ell\leftrightarrow\ell') \\
\nonumber +\frac{4 \ \delta_{i_z,j_z}}{4\pi} & \int \dd V \ 2 \ I_1^{1}(k_\ell|z) \ I_1^{1}(k_{\ell'}|z) \ I_2^{\Sigma_2}(k_{\ell},k_{\ell'}|z) \\
& \times P(k_\ell|z) \ P(k_{\ell'}|z) \ \Bigg/ \ N_\mr{gal}(i_z)^4 \ ,
\ea
and the 4-halo term from third order contributions,
\ba\label{Eq:4halo3}
\nonumber \mathcal{C}_{\ell,\ell'}^\mr{4h-3} = & \frac{2 \ \delta_{i_z,j_z}}{4\pi} \int \dd V \ 3! \ \left(I_1^1(k_{\ell},z)\right)^2 \ I_1^1(k_{\ell'},z) \ I_1^{\Sigma_3}(k_{\ell'},z) \\
& \times \ P(k_{\ell}|z) \ P(k_{\ell}|z) \ P(k_{\ell'}|z) \Bigg/ \ N_\mr{gal}(i_z)^4 \quad + \quad (\ell\leftrightarrow\ell') \ .
\ea

Finally, the most complicated case is braiding covariance, whose projection in spherical harmonics is found in \cite{Lacasa2018b}. It has some similarities with SSC in that it is also a class of terms grouped together and it also takes the form of a double redshift integral with the non-linear physics encapsulated in separable elements:
\ba\label{Eq:BraidingCov}
\nonumber \mathcal{C}_{\ell,\ell'}^\mr{braid} = 2 & \int \dd V_{ab} \ \Psi^\mr{alt}_{\ell,\ell'}(z_a) \ \Psi^\mr{alt}_{\ell,\ell'}(z_b) \\
& \times \ \mathcal{B}_{\ell,\ell'}(z_a,z_b) \Bigg/ \ N_\mr{gal}(i_z)^2 \, N_\mr{gal}(j_z)^2 ,
\ea
where
\ba\label{Eq:Braiding-kernel}
\mathcal{B}_{\ell,\ell'}(z_a,z_b) = \sum_{\ell_a} \frac{2\ell_a+1}{4\pi} \  {\threeJz{\ell}{\ell'}{\ell_a}}^2 \ C_{\ell_a}^m(z_a,z_b)
\ea
is the braiding kernel and
\ba
\Psi^\mr{alt}_{\ell,\ell'}(z) = \Big[ 2 \ I_1^{\Sigma_2}(k_{\ell'}|z) \ I_1^1(k_{\ell}|z) \, P(k_{\ell}|z) + (\ell \leftrightarrow\ell') \Big] + I_2^1(k_{\ell},k_{\ell'}|z)
\ea
encapsulates the non-linear physics.

\subsection{An approximation to braiding covariance}\label{Sect:braiding-approx}

Directly implementing Equation \ref{Eq:BraidingCov} for braiding covariance is numerically challenging. Indeed, it would need the computation of $\mathcal{B}_{\ell,\ell'}(z_a,z_b)$ for all pairs of multipoles and all pairs of redshifts. $\mathcal{B}_{\ell,\ell'}(z_a,z_b),$  itself a sum over $\mathcal{O}(\ell_\mr{max})$ multipoles, quickly makes it a burden for next-gen galaxy surveys where we target $\ell_\mr{max}=\mathcal{O}(10^3)$.

To overcome this, I devised an approximation with an approach similar to that followed by \cite{Lacasa2019} for super-sample covariance: we can approximate that $\Psi^\mr{alt,clust}_{\ell,\ell'}\Big/ \nbargal(z)^2$ varies slowly with redshift compared to $\mathcal{B}_{\ell,\ell'}$. Then
\ba
\mathcal{C}_{\ell,\ell'}^\mr{braid} \approx 2 \ \Psi^\mr{alt,int}_{\ell,\ell'}(i_z) \ \Psi^\mr{alt,int}_{\ell,\ell'}(j_z) \ B_{\ell,\ell'}(i_z,j_z),
\ea
where
\ba
\Psi^\mr{alt,int}_{\ell,\ell'}(i_z) = \int \dd V \ \Psi^\mr{alt}_{\ell,\ell'}(z)
\ea
and 
\ba
\nonumber B_{\ell,\ell'}(i_z,j_z) =& \int \dd V_{ab} \ \nbargal(z_a)^2 \, \nbargal(z_b)^2 \ \mathcal{B}_{\ell,\ell'}(z_a,z_b)  \Bigg/ \left(I^{n_g^2}(i_z) \, I^{n_g^2}(j_z)\right)\\
=& \sum_{\ell_a} \frac{2\ell_a+1}{4\pi} \  {\threeJz{\ell}{\ell'}{\ell_a}}^2 \ C_{\ell_a}^{n_g^2}(i_z,j_z)
\ea
with
\ba
I^{n_g^2}(i_z) = \int_{z\in i_z} \dd V \ \nbargal(z)^2
\ea
and
\ba
C_{\ell}^{n_g^2}(i_z,j_z) = \int \dd V_{ab} \ \nbargal(z_a)^2 \, \nbargal(z_b)^2 \ C_{\ell}^{m}(i_z,j_z) \Bigg/ \left(I^{n_g^2}(i_z) \, I^{n_g^2}(j_z)\right) .
\ea

I call this the `Bij approximation' for Braiding covariance, similarly to the name `Sij approximation' for super-sample covariance. The fact that the Sij approximation works very well \citep[see][]{Lacasa2019} proves that the Bij should work equally well, if not better. Indeed, the similarity between the separable elements $\Psi^\mr{sqz}$ and $\Psi^\mr{alt}$ \footnote{In fact, we have the reduction $\Psi^\mr{sqz}_\ell=\Psi^\mr{alt}_{\ell,\ell}$ on the diagonal.} and the fact that $\mathcal{B}_{0,0}(z,z')=\sigma^2(z,z')$ shows that $\mathcal{B}_{\ell,\ell'}$ varies quickly enough with redshift for the Bij approximation to work at $\ell=\ell'=0$. And at higher multipoles, $\mathcal{B}_{\ell,\ell'}$ only varies more quickly, making the approximation increasingly more precise. Indeed, from Eq.~\ref{Eq:Braiding-kernel}, at high $(\ell,\ell')$ $\mathcal{B}_{\ell,\ell'}$ gets contributions from $C_{\ell_a}^m(z,z')$ at high $\ell_a$ , which gets increasingly close to a Dirac $\delta(z,z')$ due to Limber approximation. These analytical arguments ensure that the Bij approximation for Braiding covariance works at least as well as the Sij approximation for SSC.

\section{Covariance results and the importance of braiding for positive definiteness}\label{Sect:cov-results}

In this section, I first present the physical and technical assumptions I used for the computation of the galaxy angular power spectrum and its covariance terms, along with the numerical results for the covariances. Then I show why these results prove the importance of including some of the non-Gaussian terms presented in Sect.~\ref{Sect:analytical-cov}: braiding and 2h1+3. Finally, I present the impact of NG terms on the measurement signal to noise ratio of the galaxy angular power spectrum.

\subsection{Setup and covariances}\label{Sect:cov-setup}

For the numerical results presented in this and later sections, I used a flat $\Lambda$CDM cosmological model with \textit{Planck} 2018 \citep{Planck2018-params} cosmological parameters $(\Omega_b h^2,\Omega_c h^2,H_0,n_S,\sigma_8)=(0.022,0.12,67,0.96,0.81)$. For the halo model, I adopted the \cite{Tinker2008} halo mass function with the corresponding halo bias from \cite{Tinker2010}. For the HOD, I used one similar to \cite{Zehavi2011}: $N_\mr{gal}=N_\mr{cen}+N_\mr{sat}$, with a Bernoulli distribution for the central galaxy with probability,
\ba
P(N_\mr{cen}=1) = \frac{1}{2} \left(1 + \mr{Erf}\left(\frac{\log_{10}M-\log_{10}M_\mr{min}}{\sigma_{\mr{log}M}}\right)\right)
\ea
and a Poisson distribution for the satellite galaxies, conditioned to the presence of the central, with mean,
\ba
\mathbb{E}\left[N_\mr{sat}|N_\mr{cen}=1\right]= \left(\frac{M}{M_\mr{sat}}\right)^{\alpha_\mr{sat}} .
\ea
In this section, I consider a single redshift bin for the galaxies: $0.9<z<1.019$. For the HOD parameters, I used $\log_{10}M_\mr{min}=11.3$, $\sigma_{\mr{log}M}=0.5$, $M_\mr{sat} = 10 \times M_\mr{sat}$ and $\alpha_\mr{sat}=1$. These parameters predict a galaxy density at these redshifts equal to the predicted one for the \Euclid{} photometric sample, that is, 3 galaxies/arcmin$^2$ (see Appendix~\ref{App:HOD}) which corresponds to a total of $\sim$450M  galaxies as I assume a full sky setup.

\begin{figure}[!ht]
\begin{center}
\includegraphics[width=\linewidth]{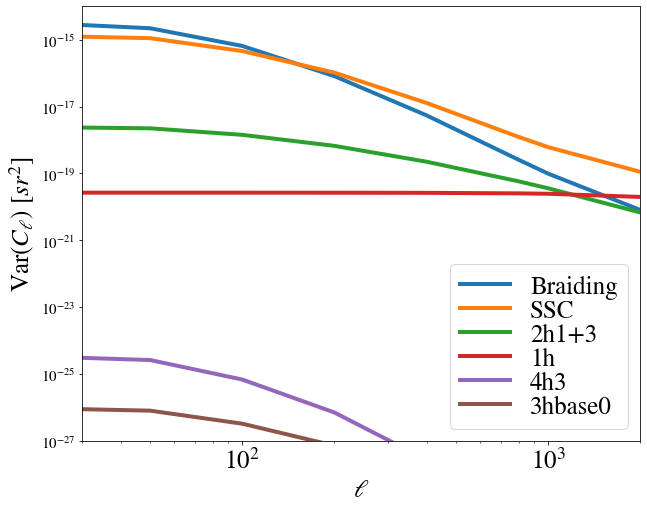}
\caption{Different non-Gaussian contributions to the variance of the angular power spectrum per multipole in the redshift bin $0.9<z<1.019$.}
\label{Fig:Var-Cl}
\end{center}
\end{figure}

With these parameters, I computed the galaxy angular power spectrum $\Clgal$ and the different non-Gaussian covariance contributions listed in Sect.~\ref{Sect:analytical-cov} for nine individual multipoles distributed logarithmically in [30,3000]. The variance per multipole created by each term is shown in Fig.~\ref{Fig:Var-Cl} plotted as a function of multipole $\ell$.

We first see that the 3h-base0 and 4h-3 terms are negligible compared to all other terms. This means that the perturbative contributions to variances are excellently encapsulated inside super-sample covariance and braiding covariance. We can then focus on the other covariance terms considered in this article: braiding and 2h1+3. We see that braiding is actually the dominant NG contribution to the variance on large scales and remains non-negligible on most of the multipole range. The 2h1+3 term is subdominant everywhere, but it still is not negligible. I emphasise that these results are not enough to draw conclusions on the importance of the terms as they only show the diagonal, rather than the whole structure of the covariance matrices.

To examine the covariance matrices and be more representative of a survey analysis, I needed to consider not only a few multipoles but the full multipole range. Computing the covariance matrices for all single multipoles in this range is not desirable, however, because (i) it is very intensive numerically and (ii) it would not be representative of actual data analysis that bins multipoles together.
Hence, I  performed a binning of multipoles, which consisted of interpolating and binning from the nine original multipoles to 29 bins distributed logarithmically $\Delta\ell/\ell = \mr{cst}$ in the range $\ell\in[32,2290]$. Hereafter, binned quantities are plotted with the indication of the central multipole of the bin, defined as the geometrical average of the bin stakes.

\begin{figure}[!ht]
\begin{center}
\includegraphics[width=\linewidth]{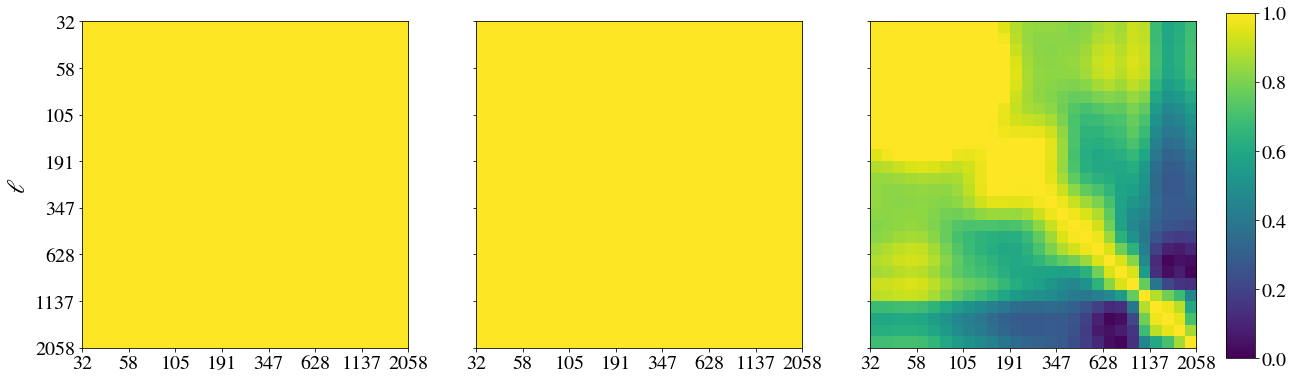}
\includegraphics[width=\linewidth]{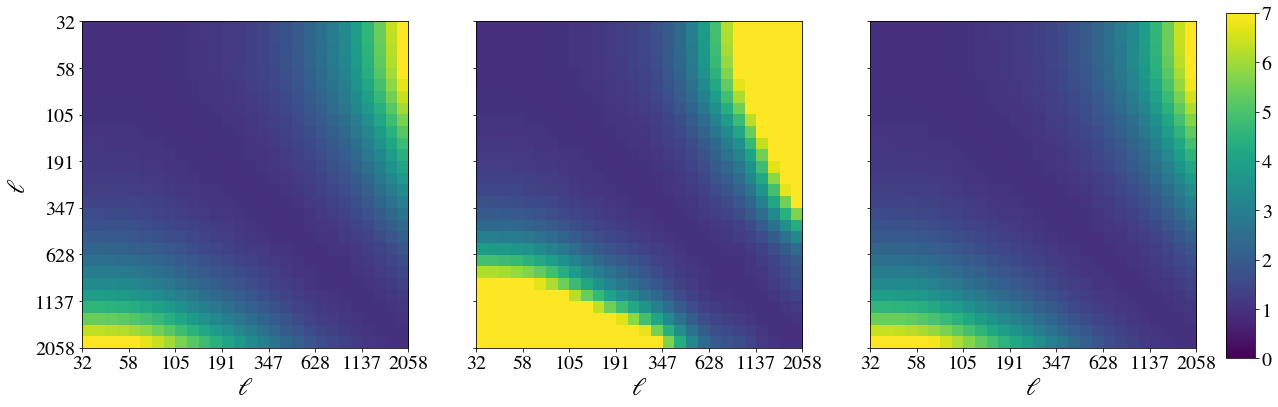}
\caption{Correlation matrices for the different non-Gaussian covariance terms, normalised by its own diagonal. \textit{Top:} SSC, 1h, Braiding. \textit{Bottom:} 2h1+3, 3h-base0, 4h-3; the color bar is clipped at 7.}
\label{Fig:Corr-Cl}
\end{center}
\end{figure}

With these specifications, I show in Fig. \ref{Fig:Corr-Cl} the correlation matrices: $C_{i,j}/\sqrt{C_{i,i} C_{j,j}}$ for each of the non-Gaussian covariance terms. Each term is normalised by its own diagonal to reveal its specific structure. I note that this is different from the more customary normalisation by the total diagonal, which lets us appreciate the relevance of the terms; however, this relevance will be addressed later, in Sect.~\ref{Sect:SNR} and \ref{Sect:impact-constraints}.

In the top row we see well-behaved terms which yield matrices with all eigenvalues $\geq 0$ : SSC, 1-halo and  braiding. The correlation coefficients are all in [-1,1]. In the bottom row we see the 2h1+3, 3h-base0 and 4h-3 terms for which the correlation coefficients can be >1 (up to 7.4 for 2h1+3, 39 for 3h-base0 and 7.5 for 4h-3; the color bar is clipped to 7 in the plots for readability), indicating that these matrices have negative eigenvalues.

\subsection{Importance of braiding for positive definiteness}\label{Sect:braiding-necessity}

In this section I examine the problem of the NG terms with negative eigenvalues: 2h1+3, 3h-base0, and 4h-3. I first give an analytical explanation why they yield, alone, correlation coefficients >1, then I give a physical explanation why they cannot be included alone and argue why Braiding covariance is necessary to regulate them to obtain a well-behaved total covariance matrix, that is, positive definite.

First, let us become convinced in an analytical sense that the correlation coefficients >1 seen in the bottom row of Fig.~\ref{Fig:Corr-Cl} are physical and not a bug in my computation. For this, I focus on the case of the 2h1+3 term. Both for simplicity, so as not to repeat similar computations thrice, and because it dominates the 3h-base0 and 4h-3 terms as seen in Fig.~\ref{Fig:Var-Cl}.\\
Let us evaluate the correlation coefficient
\be
r_{\ell,\ell'} = \frac{\mathcal{C}_{\ell,\ell'}}{\sqrt{\mathcal{C}_{\ell,\ell} \times \mathcal{C}_{\ell',\ell'}}}
\ee
for the 2h1+3 term Eq.~\ref{Eq:2halo1+3} in the following case: infinitesimally small redshift bins and $k_\ell , k_{\ell'} \ll 1/R$, where $R$ is the typical radius of a halo, so that $u(k)\rightarrow 1$. These conditions mean that the redshift integrals can be replaced by a multiplication with $\Delta z$ (which vanishes in the ratio) and that all halo model integrals $I_\mu^\beta$ are independent of $\ell,\ell'$. Then we get 
\ba
\left(r_{\ell,\ell'}^\mr{2h1+3}\right)^2 \approx & \frac{\left(I_1^1 I_3^1 P(k_\ell)+ (\ell \leftrightarrow \ell')\right)^2}{2 I_1^1 I_3^1 P(k_\ell) \times 2 I_1^1 I_3^1 P(k_{\ell'})} = \frac{1}{4}\frac{\left(P(k_\ell)+P(k_{\ell'})\right)^2}{P(k_\ell) \, P(k_{\ell'})} .
\ea
Now I further take the condition $k_\mr{eq} < k_\ell \ll k_{\ell'}$, where $k_\mr{eq}$ is the position of the maximum of the matter power spectrum $P(k)$ (corresponding to matter-radiation equality) so that both wave vectors are in the decreasing part of $P(k)$. In that case $P(k_\ell) \gg P(k_{\ell'})$ and we get
\ba
r_{\ell,\ell'}^\mr{2h1+3} \approx & \frac{1}{2}\sqrt{\frac{P(k_\ell)}{P(k_{\ell'})}} > 1 .
\ea
So the result is physical: alone these covariance terms give correlation coefficients which can be >1. This means that these terms yield incorrect covariance matrices if left alone: two measurements can be more than 100\% correlated, or in other term the matrix restricted to these two points has a negative eigenvalue.

\begin{figure}[!ht]
\begin{center}
\includegraphics[width=\linewidth]{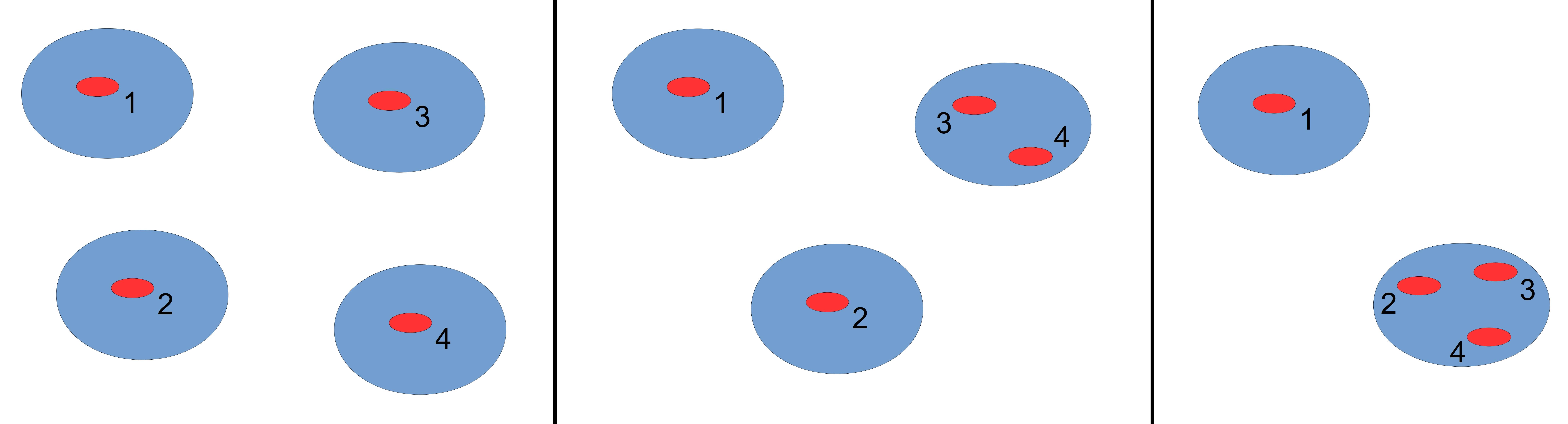}
\caption{Diagrams for some of the trispectrum terms involved in the covariance of the galaxy angular power spectrum $\Cov\left(C_\ell^\mr{gal},C_{\ell'}^\mr{gal}\right)$. \textit{From left to right:} 4-halo, 3-halo, and 2-halo 1+3 term. Galaxies 1 and 2 are the source of the first power spectrum $C_\ell^\mr{gal}$, while galaxies 3 and 4 are the source of the second power spectrum $C_{\ell'}^\mr{gal}$}
\label{Fig:Diagrams-4h3h2h1p3}
\end{center}
\end{figure}

This result can also be understood more visually by using the diagrammatic formalism built by \cite{Lacasa2014}. As shown by \cite{Lacasa2018b}, the 4h-3 is part of the terms of the left diagram of Fig.~\ref{Fig:Diagrams-4h3h2h1p3}, which quantifies how the 2-halo part of the spectrum is correlated with itself due to halos being clustered in a (non-Gaussian) matter field. The 3h-base0 is part of the terms of the central diagram, which quantifies how the 2-halo part of the spectrum is correlated with the 1-halo part due to halos being clustered in a (non-Gaussian) matter field. And the 2h1+3 is the entirety of the terms of the right diagram, which quantifies how the 2-halo part of the spectrum is correlated with the 1-halo part due to halo coincidence. From these diagrams it becomes clear that the 2h1+3 term is going to be maximal when $\ell$ is in the large-scale 2-halo dominated regime while $\ell'$ is in the small-scale 1-halo dominated regime. So this term is going to yield high covariance when $\ell' \gg \ell$ and minimal covariance when $\ell=\ell'$, i.e. exactly the off-diagonal behaviour we see in Fig.~\ref{Fig:Corr-Cl}.

Now this behaviour has to be regulated by another covariance term which makes the total covariance matrix well-behaved. Mathematically, the regulator cannot be the Gaussian part of the covariance, nor SSC, nor the 1h trispectrum term alone. First, it cannot be the Gaussian part of the covariance. Indeed, going to arbitrarily high redshifts, we can have arbitrarily high multipoles $\ell'$ that fulfill the conditions $k_{\ell'}\sim \ell'/r(z) \ll 1/R$. At these multipoles, the Gaussian variance becomes negligible since it decreases as $1/(2\ell'+1)$. Second, this cannot either be the super-sample covariance. Indeed, SSC gives a near degenerate covariance matrix with a single positive eigenvalue, the other being zero, as seen from Fig.~\ref{Fig:Corr-Cl} where the correlation matrix is 100\% everywhere. So SSC cannot regulate a multitude of negative eigenvalues. Finally, for the same reason, the regulator cannot either be the 1h trispectrum term, which is constant on large scales.

\begin{figure}[!ht]
\begin{center}
\includegraphics[width=\linewidth]{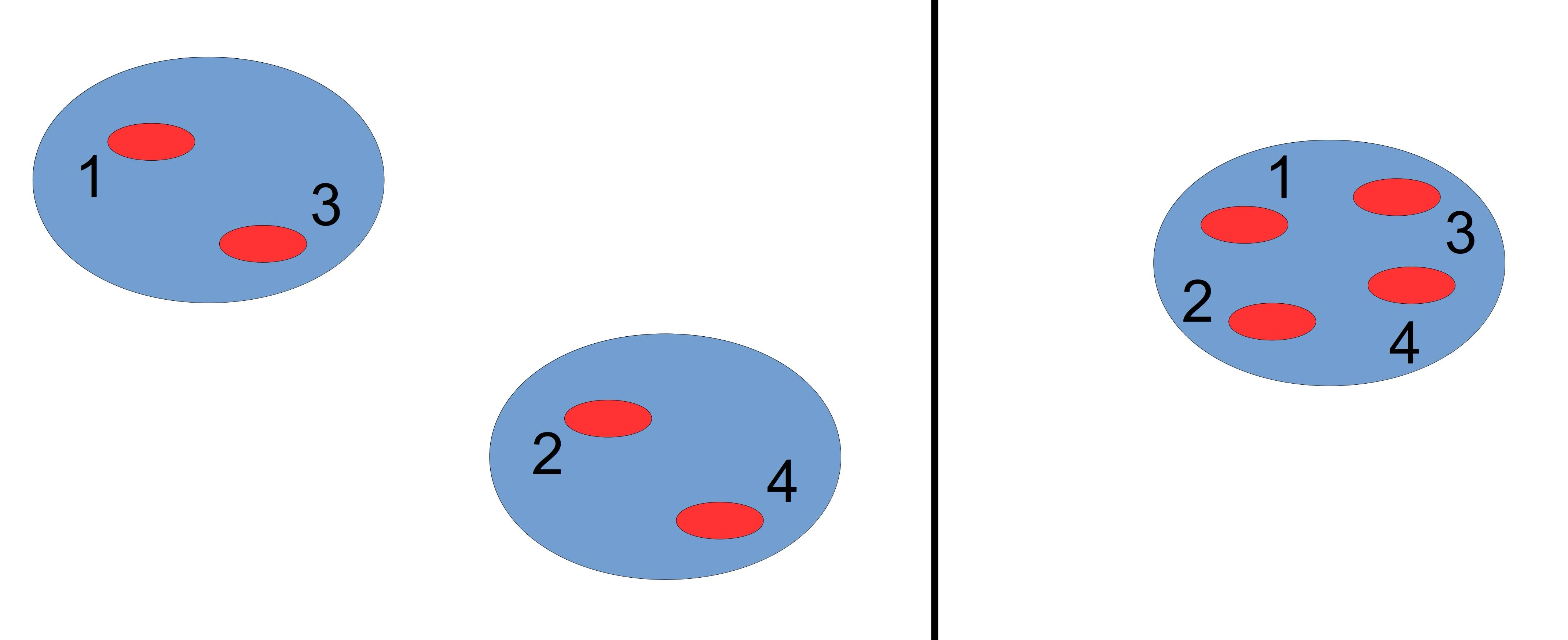}
\caption{Diagrams for some of the trispectrum terms involved in the covariance of the galaxy angular power spectrum. \textit{Left:} 2-halo part of Braiding, \textit{right:} 1-halo term.}
\label{Fig:Diagrams-Braid1h}
\end{center}
\end{figure}

We can find the regulator via the diagram discussion. Since the 2h1+3 term quantifies how the 2-halo part of the spectrum is correlated with the 1-halo part due to halo coincidence, it has to be regulated by a first term which quantifies how the 2-halo part of the spectrum is correlated with itself due to halo coincidence, and a second term which quantifies how the 1-halo part of the spectrum is correlated with itself due to halo coincidence. The first wanted term is part of braiding covariance: it is the 2-halo part of Braiding, which corresponds to the left diagram of Fig.~\ref{Fig:Diagrams-Braid1h}. The second wanted term is the 1-halo trispectrum term, which corresponds to the right diagram of Fig.~\ref{Fig:Diagrams-Braid1h}.

\begin{figure}[!ht]
\begin{center}
\includegraphics[width=.7\linewidth]{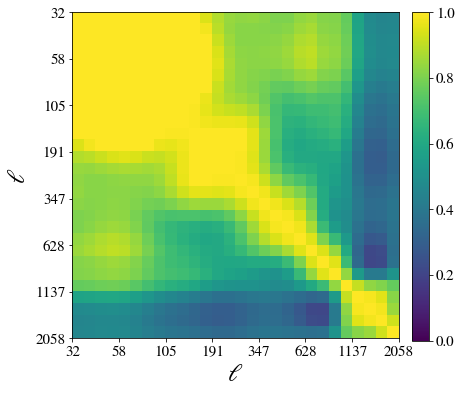}
\caption{Correlation matrix for ONG, the sum of non-Gaussian covariance terms other than SSC: 1h, 2h1+3, 3h-base0, 4h-3, and Braiding.}
\label{Fig:Corr-Cl-ONG}
\end{center}
\end{figure}

With similar considerations, we can see that the regulator of the 3h-base0 and 4h-3 terms is Braiding covariance. So its is the sum of the 1h, Braiding, 2h1+3, 3h-base0, and 4h-3 terms that yield a well-behaved covariance. In the following I call this sum `other non-Gaussianity' (ONG) by contrast with the non-Gaussian covariance that has been the most studied to date: super-sample covariance. For comparison, in the previous literature non-SSC NG terms have also been called `in-survey' \citep[e.g.][]{Rizzato2019}, `connected non-Gaussian (cNG)' \citep[e.g.][]{Barreira2018}, `trispectrum' \citep[e.g.][]{Li2014} or `T0' \citep[e.g.][]{Wadekar2019}. Figure~\ref{Fig:Corr-Cl-ONG} shows the correlation matrix for the ONG group.

We see that the ONG indeed has all correlation coefficient $\leq$ 100\%. Furthermore, numerical investigation shows that all eigenvalues are >0. Thus, the addition of braiding covariance has correctly regulated the off-diagonal components of the 2h1+3, 3h-base0, and 4h-3 terms. I conclude that the inclusion of Braiding is necessary to go beyond the current state of the art for non-Gaussian covariances.

\subsection{Impact on the signal-to-noise ratio}\label{Sect:SNR}

\begin{figure}[!ht]
\begin{center}
\includegraphics[width=\linewidth]{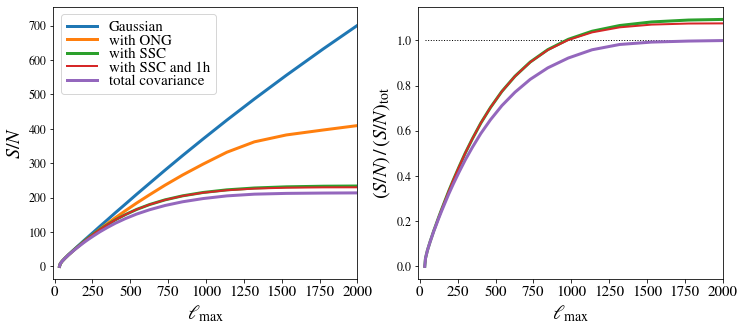}
\caption{Cumulative signal to noise ratio for the measurement of $\Clgal$ in the bin $0.9<z<1.019$ as a function of maximum multipole of analysis. \textit{Left, from top to bottom:} Gaussian covariance only, Gaussian + `other NG', Gaussian + SSC, Gaussian + SSC + 1h, total covariance. \textit{Right:} zoom on the three lowest curves: Gaussian + SSC, Gaussian + SSC + 1h, total covariance, all normalised by the value of the signal to noise using the full multipole range and the total covariance.}
\label{Fig:SNR-Cl}
\end{center}
\end{figure}

Alhough Braiding is necessary to include ONG covariance, the question remains of whether ONG has a significant impact on the information content of the galaxy angular power spectrum. In this section, I use the signal-to-noise ratio (S/N)\ba
\left(\frac{S}{N}\right)^2 = \sum_{\ell,\ell'=\ell_\mr{min}}^{\ell_\mr{max}} C_\ell^\mr{gal} \ \mathcal{C}^{-1}_{\ell,\ell'} \ C_{\ell'}^\mr{gal}
\ea
as a first metric to quantify this information content, as already used in the literature \citep[e.g.][]{Rizzato2019}.\\
To this end, I also use the halo model at tree-level to predict the power spectrum $C_{\ell'}^\mr{gal}$. This modelling allows for $\sim 10\%$ precision; for future surveys, this is sufficient for the prediction of the covariance, but not for the prediction of the power spectrum. This is, however, not an issue for this analysis as my goal here is to gauge the relative impact of covariance terms.
Figure~\ref{Fig:SNR-Cl} shows $S/N$ plotted as a function of $\ell_\mr{max}$
for different degree of sophistication in the computation of the covariance.

If the analysis is carried out on the full range of multipoles, as is scheduled, for instance, for \Euclid, then non-Gaussian covariance terms have a large impact on the information content. Compared to the Gaussian case, ONG alone decreases $S/N$ by a factor 1.7. This is clearly a large impact, and one must go beyond Gaussian covariances. Now the current state of the art includes super-sample covariance, and that term has a larger impact: SSC alone decreases $S/N$ by a factor 3.1. Finally, when accounting for the total covariance: Gaussian+SSC+ONG, $S/N$ decreases by a factor 3.4 compared to the Gaussian case. So ONG has a 9.4\% impact on top of SSC. The 1h covariance has a negligible impact on top of SSC, so the bulk of the 9.4\% impact comes from the Braiding and 2h1+3 terms.

Thus, including ONG seems fairly important (if SSC is already accounted for) given, for example, that \Euclid{} has a requirement of 10\% precision on error bars. First, I argue that ONG should still be accounted for because it makes the information systematically lower and, thus, error bars become systematically larger. Second, this section used the $S/N$ in a single redshift bin as a metric and the question remains open of the impact on parameter constraints when summed over the entire redshift range. This is the subject of the next section.

\section{Impact on parameter constraints}\label{Sect:impact-constraints}

\subsection{Setup}\label{Sect:impact-setup}

I use survey specifications representative of the \Euclid{} photometric galaxy sample \citep{Euclid-IST}: sky coverage $f_\mr{SKY}=0.36$, a galaxy redshift distribution of
\ba\label{Eq:Ngal(z)-Euclid}
n(z) \propto  \left( \frac{z}{z_0} \right)^2 \exp{\left[ - \left( \frac{z}{z_0} \right)^{3/2} \right],} 
\ea
where $z_0 = z_{\rm m}/\sqrt{2,}$ with $z_{\rm m}= 0.9$ the median redshift \citep{Euclid-redbook}. The  total density is 30 gals$\cdot \mr{arcmin}^{-2}$ in the redshift range [0,2.5]. Following \cite{Euclid-IST}, the sample is divided into 10 equi-populated redshift bins, whose bin stakes are $z={0.001,0.418,0.56,0.678,0.789,0.9,1.019,1.155,1.324,1.576,2.5}$ \footnote{Hence the bin $0.9<z<1.019$ considered in Sect.~\ref{Sect:cov-results} is the 6th bin in the analysis of this section.}. To reproduce this redshift distribution with the halo model, I use the Halo Occupation Distribution described in Sect.~\ref{Sect:cov-setup}, further including a redshift dependence of $M_\mr{min}$ in the form:
\ba
M_\mr{min}(z) = M_\mr{min}^a + M_\mr{min}^b \, z + M_\mr{min}^c \, z^2 + M_\mr{min}^d \, z^3 .
\ea
As shown in Appendix \ref{App:HOD}, this parametrisation allows to reproduce the \Euclid{}-expected galaxy counts to 2.5\% precision, and predicts a galaxy bias consistent with simulations.

In this section, I quantify the impact of covariances on parameter constraints using the methodology of Fisher forecasts. To this end, I use both Fisher matrices in a given redshift bin:
\ba
F_{\alpha,\beta}(i_z) = \sum_{\ell,\ell'=\ell_\mr{min}}^{\ell_\mr{max}} \partial_\alpha C_\ell^\mr{gal}(i_z) \ \mathcal{C}^{-1}_{\ell,\ell'}(i_z,i_z) \ \partial_\beta C_{\ell'}^\mr{gal}(i_z)
\ea
and summed over all bins:
\ba
F_{\alpha,\beta} = \sum_{i_z,j_z} \sum_{\ell,\ell'=\ell_\mr{min}}^{\ell_\mr{max}} \partial_\alpha C_\ell^\mr{gal}(i_z) \ \mathcal{C}^{-1}_{\ell,\ell'}(i_z,j_z) \ \partial_\beta C_{\ell'}^\mr{gal}(j_z),
\ea
where $\alpha,\beta$ are model parameters, that is, cosmological and HOD parameters in the following : $(\Omega_b h^2, \Omega_c h^2, H_0, n_S, \sigma_8, w_0)$ and $(\alpha_\mr{sat}, \sigma_{\log M}, M_\mr{ratio}, M_\mr{min}^a, M_\mr{min}^b, M_\mr{min}^c, M_\mr{min}^d)$; $\partial_\alpha$ is the derivative of the observable w.r.t. parameter $\alpha$.

\subsection{Impact on cosmological parameters}\label{Sect:impact-cosmo}

\begin{figure}[!ht]
\begin{center}
\includegraphics[width=\linewidth]{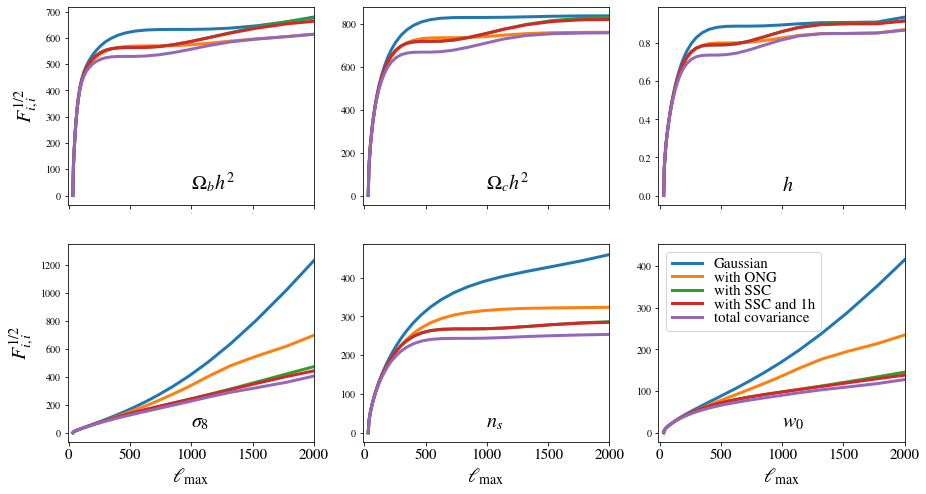}
\caption{(Square root of the) cumulative Fisher elements for the cosmological parameters in the considered redshift bin, as a function of the maximum multipole of the analysis.}
\label{Fig:Fisher-cosmo}
\end{center}
\end{figure}

We can first look at the Fisher matrix elements in a given redshift bin. For the purposes of illustration, I chose the bin $0.9<z<1.019$, which is the same bin as in Sect.~\ref{Sect:cov-results}, containing the median redshift of the galaxy sample and whose results I found representative of the whole sample. Figure~\ref{Fig:Fisher-cosmo} shows, as a function of the maximum multipole of analysis $\ell_\mr{max}$, the square root of the Fisher elements for each cosmological parameter of the $w$CDM model. This quantity is the inverse of the error bar on the considered parameter if all other (cosmological and HOD) parameters were perfectly known.

If the analysis is carried out on the full range of multipoles then non-Gaussian covariance terms would have a mild impact on the information content for the three first parameters: $\Omega_b h^2$, $\Omega_c h^2,$ and $h$, with ONG being more significant than SSC. By contrast, non-Gaussian terms have a large impact on the three last parameters: $\sigma_8$, $n_S,$ and $w_0$. These three latter parameters are arguably the most interesting to constrain with surveys of the large scale structure. The measurement of $\sigma_8$ is interesting in the context of the current tension between local measurements and the CMB. The parameter $n_S$ helps to constrain inflation and can be seen as representative of parameters in a more extended model that would change the shape of the power spectrum, for example, a running of the spectral index or massive neutrinos. Finally, the equation of state of dark energy is one of the main science drivers of current and future galaxy surveys.

Compared to the Gaussian Fisher matrix, ONG alone decreases the Fisher content on dark energy $F^{1/2}_{w,w}$ by a factor 1.8 ; for other parameters, the factor ranges between 1.08 (for $h$) and 1.8 (for $\sigma_8$). Super-sample covariance decreases the information on dark energy by a factor 2.9; for other parameters the factor ranges between 1.01 (for $\Omega_b h^2$) and 2.6 (for $\sigma_8$). The total NG decreases $F^{1/2}_{w,w}$ by a factor 3.3; for other parameters the factor ranges between 1.08 (for $h$) and 3.1 (for $\sigma_8$). When compared to Gaussian+SSC, ONG has a 14\% impact on $F^{1/2}_{w,w}$ ; for other parameters, the impact ranges between 5.6\% (for $h$) and 16\% (for $\sigma_8$). As in the case of Sect.~\ref{Sect:SNR}, the 1h covariance has a negligible impact on top of SSC so the bulk of the ONG impact comes from the braiding and 2h1+3 terms.

\begin{figure}[!ht]
\begin{center}
\includegraphics[width=\linewidth]{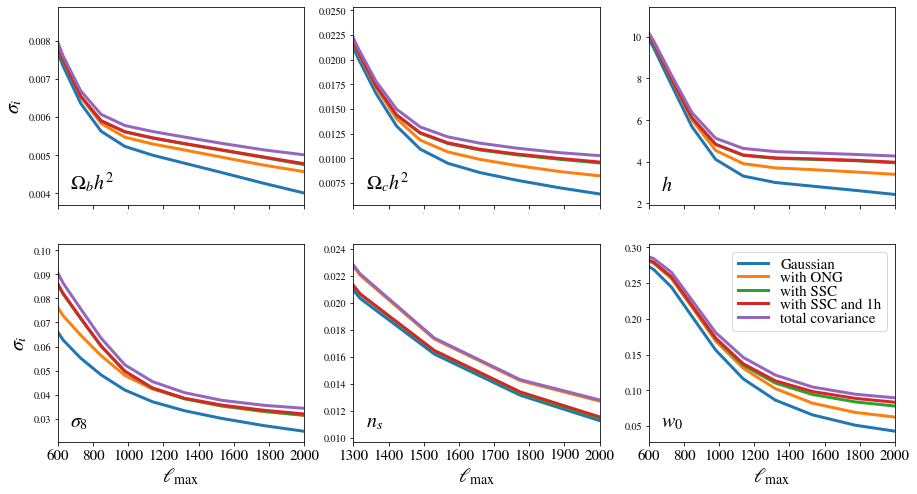}
\caption{Marginalised error bars on each cosmological parameters using all redshift bins as a function of the maximum multipole of analysis.}
\label{Fig:errorbar-cosmo}
\end{center}
\end{figure}

In a second step, I compute the Fisher matrix summed over all redshift bins. This represents the full constraining power of the mock survey; it allows for the breaking of parameter degeneracies, in particular, between parameters for the redshift dependence of the HOD which are nearly completely degenerate in a single bin. In Fig.~\ref{Fig:errorbar-cosmo} I plot the marginalised error bars $\sigma_i=\sqrt{(F^{-1})_{ii}}$ for each cosmological parameter as a function of the maximum multipole of analysis $\ell_\mr{max}$.

When using the full multipole range, non-Gaussian covariance terms have a large impact on the information content for  all cosmological parameters. Compared to the Gaussian case, ONG alone increases the error bar on $w$ by 50\%; for other parameters, the impact ranges between 14\% (for $n_S$) and 41\% (for $h$). SSC increases $\sigma_w$ by 88\%; for other parameters, the impact ranges between 1.6\% (for $n_S$) and 65\% (for $h$). The total NG increases $\sigma_w$ by 117\%; for other parameters, the impact ranges between 15\% (for $n_S$) and 79\% (for $h$). When compared to Gaussian+SSC, ONG has a 15\% impact on $\sigma_w$ ; for other parameters the impact ranges between 5.7\% (for $\Omega_b h^2$) and 13\% (for $n_S$). The ONG impact exceeds the threshold of 10\% (\Euclid{} precision requirement) for two parameters: $n_S$ and $w$ ($\sigma_8$ being affected at 9.6\%).

It is interesting to note that ONG has a larger impact than SSC on $n_S$. This happens because at first order, SSC erases information on the amplitude of the power spectrum (and the redshift dependence of this amplitude) as SSC is 100\% correlated. Once we have marginalised over $\sigma_8$, this amplitude erasing does not affect $n_S$, hence, the small (1.6\%) impact of SSC on $n_S$. By contrast, the ONG correlation matrix has a more complex structure and contains terms that couple large and small scale measurements. This affects the lever arm necessary to constrain $n_S$ more heavily. Thus, we can anticipate that other parameters which affect the shape of the matter power spectrum, such as a running of the spectral index or massive neutrinos, would also be more affected by ONG than by SSC.

\begin{figure}[!ht]
\begin{center}
\includegraphics[width=\linewidth]{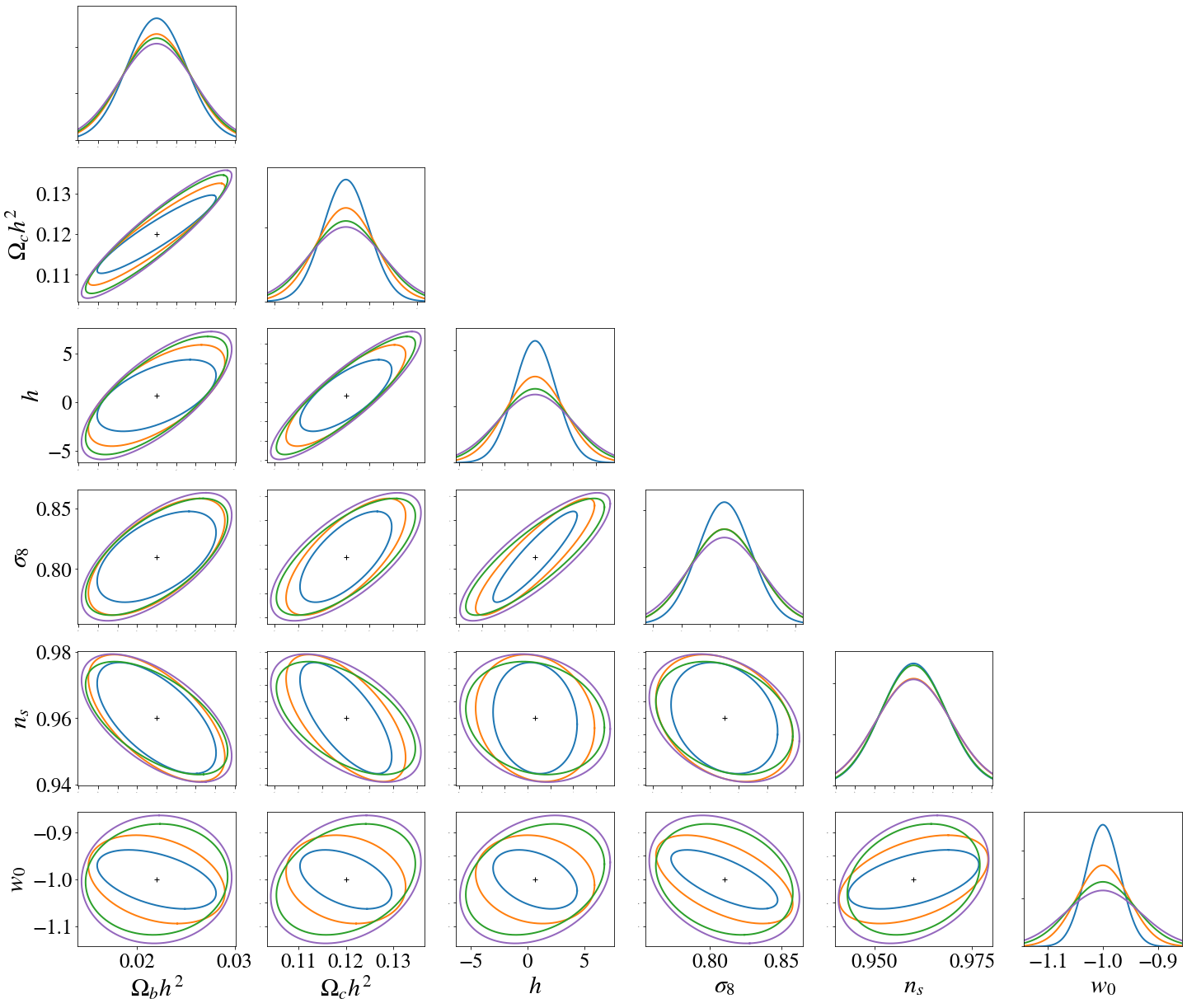}
\caption{Fisher ellipses on cosmological parameters, using all redshift bins and the full multipole range. The color coding is identical to the other figures: blue=Gaussian, orange=Gaussian+ONG, green=Gaussian+SSC, violet=total covariance.}
\label{Fig:ellipses-cosmo}
\end{center}
\end{figure}

Finally, Fig.~\ref{Fig:ellipses-cosmo} shows the Fisher plot with parameter probability distribution functions (PDFs) and 2$\sigma$ ellipses that allow for parameter degeneracies to be seen for cosmological constraints using the full multipole range and marginalised over HOD parameters with flat priors. For readability, I did not include the case of Gaussian+SSC+1h, which gives curves nearly identical to the Gaussian+SSC case.

We see that PDFs are progressively widened by non-Gaussianities. Furthermore, parameter degeneracies can be affected, sometimes in non-trivial way. For instance the direction of the degeneracy between $w$ and $\Omega_c h^2$ reverses, though the degeneracy is weak. Additionally, for the strength of the degeneracy, as evidenced by the ellipticity of the Fisher ellipses, it decreases slightly when including NG between $n_S$ and $w$, but it increases significantly between $\Omega_b h^2$, $\Omega_c h^2$ and $h$. 
This latter effect dominates the total amount of degeneracy as measured by the condition number of the Fisher matrix, which increases from $4.8\times 10^7$ in the Gaussian case to $1.4\times 10^8$ in the full non-Gaussian case. This is discussed in more details in Sect.~\ref{Sect:discu}.

\subsection{Impact on halo occupation distribution parameters}\label{Sect:impact-HOD}

We first look at the Fisher matrix elements in the redshift bin $0.9<z<1.019$. Figure~\ref{Fig:Fisher-cosmo} shows, as a function of the maximum multipole of analysis $\ell_\mr{max}$, the square root of the Fisher elements for each HOD parameter. This quantity is the inverse of the error bar on the considered parameter if all other (cosmological and HOD) parameters were perfectly known.

\begin{figure}[!ht]
\begin{center}
\includegraphics[width=\linewidth]{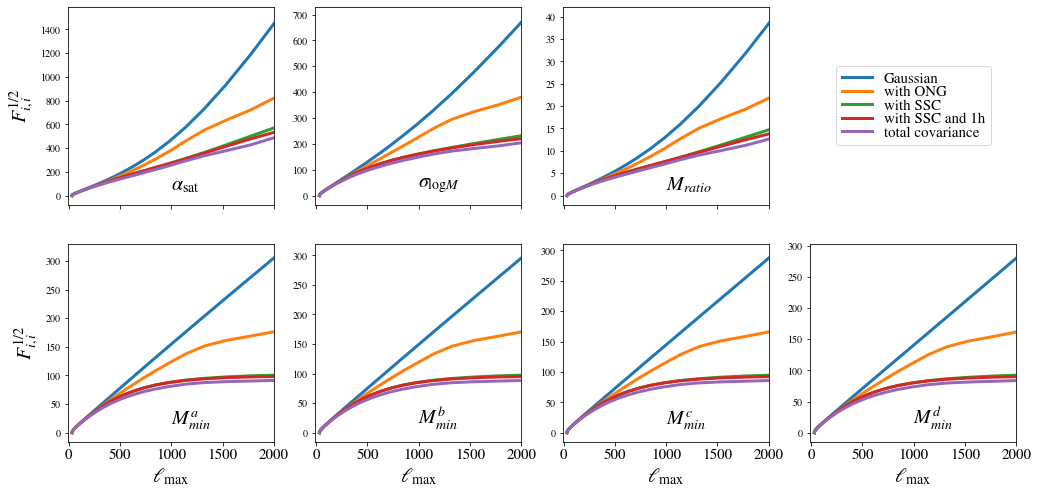}
\caption{(Square root of the) cumulative Fisher elements for the HOD parameters in the considered redshift bin as a function of the maximum multipole of analysis.}
\label{Fig:Fisher-HOD}
\end{center}
\end{figure}

If the analysis is carried out on the full range of multipoles, then non-Gaussian covariance terms have a large impact on the information content for all parameters. Compared to the Gaussian case, ONG alone decreases the Fisher content on $\alpha_\mr{sat}$, $F^{1/2}_{\alpha_\mr{sat},\alpha_\mr{sat}}$, by a factor 1.8; for other parameters, this factor is the same to the first decimal, ranging between 1.76 and 1.79. SSC decreases the information on $\alpha_\mr{sat}$ by a factor 2.6; for other parameters, the factor ranges between 2.7 (for $M_\mr{ratio}$) and 3.1 (all parameters for the redshift dependence of $M_\mr{min}$). The total NG decreases $F^{1/2}_{\alpha_\mr{sat},\alpha_\mr{sat}}$ by a factor 3; for other parameters, the factor ranges between 3.1 (for $M_\mr{ratio}$) and 3.4 (all parameters for the redshift dependence of $M_\mr{min}$). When compared to Gaussian+SSC, ONG has a 17\% impact on $F^{1/2}_{\alpha_\mr{sat},\alpha_\mr{sat}}$ ; for other parameters the impact ranges between 10\% (all parameters for the redshift dependence of $M_\mr{min}$) and 16\% (for $M_\mr{ratio}$). As in the case of cosmological parameters (Sect.~\ref{Sect:impact-cosmo}) and the $S/N$ (Sect.~\ref{Sect:SNR}), the 1h covariance has a negligible impact on top of SSC, so the bulk of the ONG impact comes from the braiding and 2h1+3 terms.

\begin{figure}[!ht]
\begin{center}
\includegraphics[width=\linewidth]{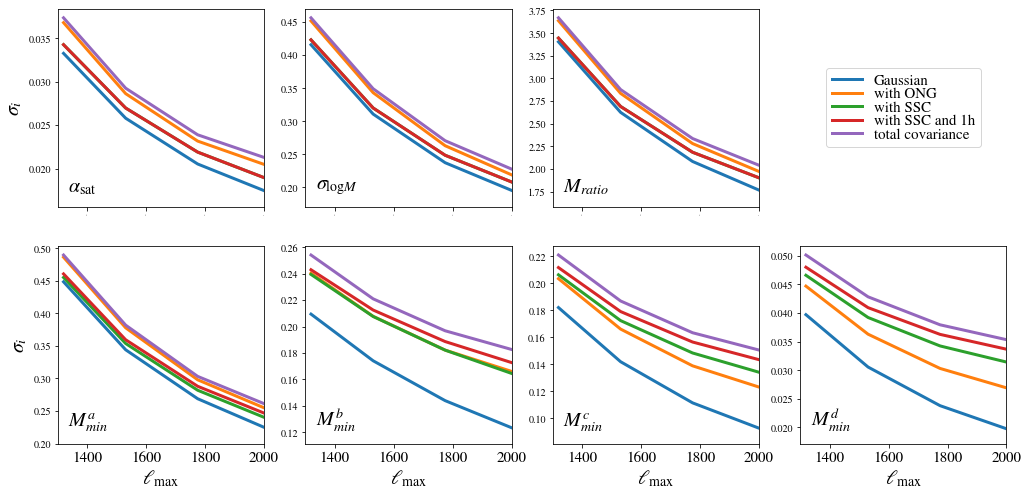}
\caption{Marginalised error bars on each HOD parameters using all redshift bins as a function of the maximum multipole of analysis.}
\label{Fig:errorbar-HOD}
\end{center}
\end{figure}

I now move to the Fisher matrix summed over all redshift bins. In Fig.~\ref{Fig:errorbar-HOD}, I plot the marginalised error bars $\sigma_i=\sqrt{(F^{-1})_{ii}}$ for each HOD parameter, as a function of the maximum multipole of analysis $\ell_\mr{max}$.

When using the full multipole range, non-Gaussian covariance terms have a large impact on the information content for \emph{all} HOD parameters. Compared to the Gaussian case, ONG alone increases the error bar on $\alpha_\mr{sat}$ by 19\% ; for other parameters, the impact ranges between 12\% (for $M_\mr{ratio}$) and 39\% (for $M_\mr{min}^d$). SSC increases $\sigma_{\alpha_\mr{sat}}$ by 9\%; for other parameters, the impact ranges between 7\% (for $\sigma_{\log M}$) and 64\% (for $M_\mr{min}^d$). The total NG increases $\sigma_{\alpha_\mr{sat}}$ by 24\%; for other parameters, the impact ranges between 17\% (for $M_\mr{ratio}$) and 85\% (for $M_\mr{min}^d$). When compared to Gaussian+SSC, ONG has a 13\% impact on $\sigma_{\alpha_\mr{sat}}$ ; for other parameters. the impact ranges between 7.4\% (for $M_\mr{ratio}$) and 13\% (for $M_\mr{min}^d$). The ONG impact is generally stronger than for cosmological parameters, exceeding the threshold of 10\% (\Euclid{} precision requirement) for four parameters: $\alpha_\mr{sat}$, $M_\mr{min}^b$, $M_\mr{min}^c$ and $M_\mr{min}^d$.

Interestingly, the impact of ONG is greater than that of SSC for four parameters: $\alpha_\mr{sat}$, $\sigma_{\log M}$, $M_\mr{ratio}$ , and $M_\mr{min}^a$. Furthermore. we can note that, for $M_\mr{min}^b$, $M_\mr{min}^c$ ,and $M_\mr{min}^d$, the inclusion of the 1h covariance makes a visible difference on top of SSC for once, although the rest of in-survey covariance and braiding are necessary to reproduce the full error bar.

\begin{figure}[!ht]
\begin{center}
\includegraphics[width=\linewidth]{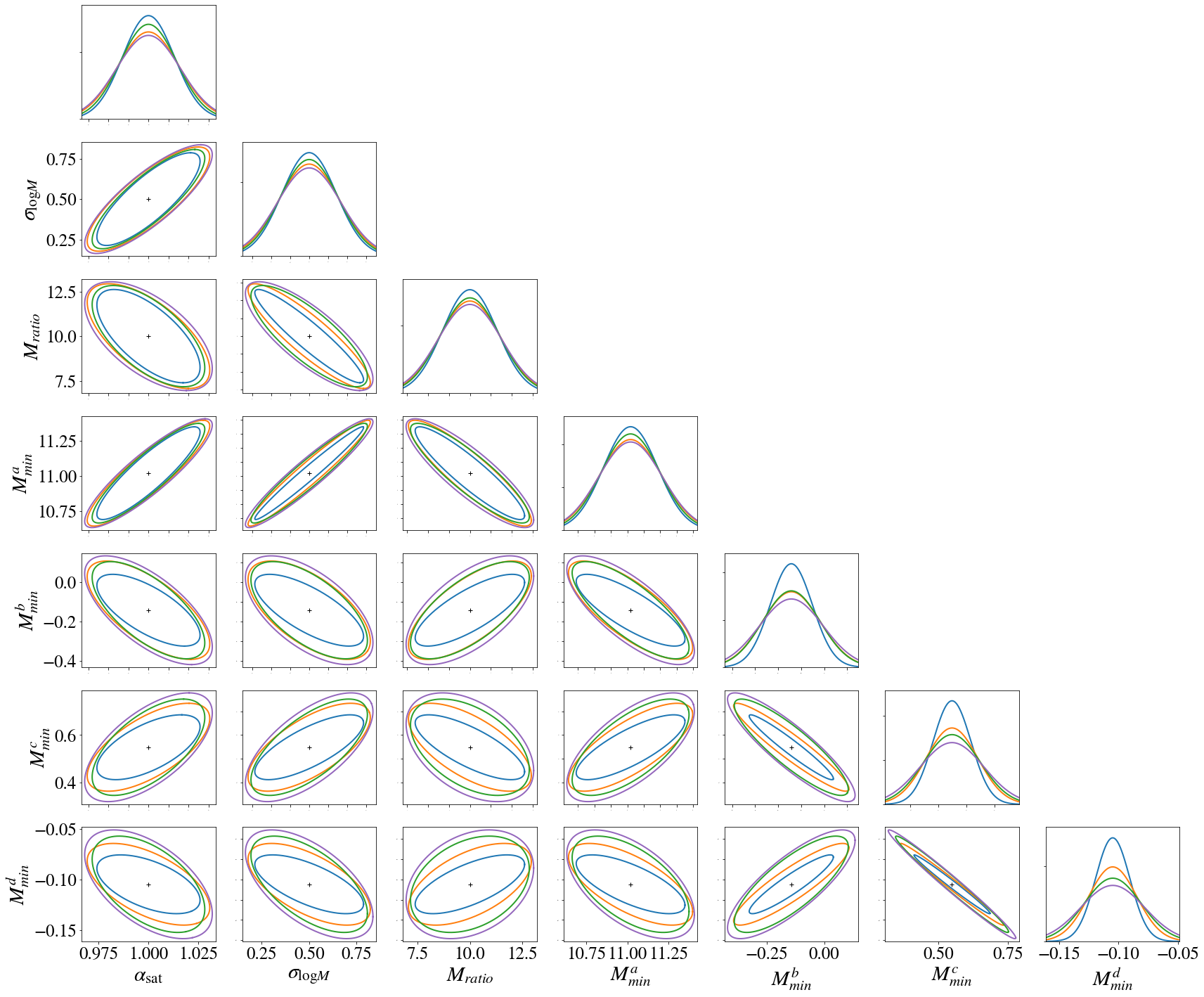}
\caption{Fisher ellipses on HOD parameters, using all redshift bins and the full multipole range. The color coding is identical to the other figures: blue=Gaussian, orange=Gaussian+ONG, green=Gaussian+SSC, violet=total covariance.}
\label{Fig:ellipses-HOD}
\end{center}
\end{figure}

Finally, Fig.~\ref{Fig:ellipses-HOD} shows the Fisher plot with parameter PDFs and 2$\sigma$ ellipses for HOD constraints using the full multipole range and marginalised over cosmological parameters with flat priors.

Again, PDFs are progressively widened by non-Gaussianities. Furthermore, the strength of parameter degeneracies is generally eased by NG. This is evidenced by the condition number of the Fisher matrix, which decreases from $5.6\times 10^7$ in the Gaussian case to $3.9\times 10^7$ in the full non-Gaussian case.

\section{Discussion}\label{Sect:discu}

By way of a summary of previous results, I have developed an implementation of non-Gaussian covariance terms for galaxy clustering that were initially derived in \cite{Lacasa2018b}. I developed a numerically tractable approximation for braiding covariance and shown that this class of terms is necessary to include other in-survey covariance terms. Grouping braiding and in-survey under the term ONG covariance, I then studied its impact on $S/N$ analysis and Fisher forecast on the $w$CDM model with the angular power spectrum with \Euclid-like galaxy specifications.

ONG by itself has a large impact on all astrophysical and cosmological parameters, ranging between 12\% and 50\%. This impact is lowered to some extent by the other NG contender: SSC, which is already included in some current analyses. Compared to this Gaussian+SSC state of the art, ONG still has a significant impact on the covariance, a result in agreement with \cite{Barreira2018b} for weak lensing ; it can even dominate SSC in some configurations, a result that is in agreement with \cite{Wadekar2019}, which appeared after the first version of this article came out. For parameter constraints, the impact on marginalised error bars ranges between 6\% and 15\% ; it exceeds 10\% --\Euclid{} precision requirement-- for the majority of HOD parameters and a couple of cosmological parameters of the $w$CDM model.\\
A parameter of particular interest is $n_S$, whose constraints are significantly affected by ONG. As SSC mostly impacts information on the power spectrum amplitude in opposition to its shape, I expect that  ONG should also affect other extensions of the standard cosmological model that change the shape of the matter power spectrum, such as massive neutrinos, warm dark matter and a running of the spectral index.

\begin{table}
\begin{center}
\begin{tabular}{c|c|c||c|c}                     
& \multicolumn{2}{|c||}{Before marginalisation} &  \multicolumn{2}{c}{After marginalisation} \\ 
\cline{2-5}
& SSC+1h & Total NG & SSC+1h & Total NG \\
\hline
$\sigma_8$ & +340\% & +360\% & +31\% & +41\% \\
\hline
$n_S$ & +70\% & +84\% & +3\% & +15\% \\
\hline
$w$ & +290\% & +310\% & +10\% & +120\% \\
\end{tabular}
\end{center}
\caption{For a few cosmological parameters, increase of the error bars compared to the Gaussian case when using the full multipole range and all redshift bins.}\label{Table:increase-errors-allz}
\end{table}

Interestingly, the increase of error bars due to NG is stronger when the other parameters are fixed, and less strong after marginalisation\footnote{I always marginalise with flat uninformative priors. If informative external priors were to be applied to some parameters, the other parameters would feel a stronger NG impact.}, as evidenced by Table~\ref{Table:increase-errors-allz}. This happens because the Gaussian Fisher matrix generally has stronger parameter degeneracies compared to the non-Gaussian covariance. The inclusion of NG often increases the minor axis of the Fisher ellipses more than the major axis, leading to a decrease of ellipticity. This is evidenced by the condition number of the whole Fisher matrix (HOD+cosmological parameters) which decreases from $1.0\times 10^9$ in the Gaussian case to $6.5\times 10^8$ in the full non-Gaussian case\footnote{This decrease is also present, and even more pronounced, if I look at the condition number for the Fisher correlation matrix $F_{ij}/\sqrt{F_{ii} F_{jj}}$}. Physically, what happens is that with a Gaussian covariance, we erroneously attribute very small error bars to the small scales; so the constraining power is located in a small number of small-scale measurements, leading to parameter degeneracies. By contrast, when NG is accounted for, error bars are increased on small scales so the constraining power is distributed more evenly among scales.

The only exceptions to this argument are $\Omega_b h^2$, $\Omega_c h^2,$ and $h,$ where the strength of degeneracies is increased by NG. First, I checked that this degeneracy is not an effect of the choice of parameters; it is still present if I use $(\Omega_b,\Omega_c,h)$ instead of $(\Omega_b h^2,\Omega_c h^2,h)$. Second, this increase of degeneracy happens because these parameters are mostly constrained by the redshift dependence of the clustering amplitude. This information is heavily affected by SSC. In terms of the likelihood approached to SSC developed in \cite{Lacasa2019}, these parameters become degenerate with the redshift evolution of the background change $\delta_b(z)$. Indeed, we see from Fig.~\ref{Fig:ellipses-cosmo} that the largest increase of the degeneracy comes from SSC.

In looking at the condition numbers, I find that it is worsened by NG for cosmological parameters ($4.8 \times 10^7 \rightarrow 1.4 \times 10^8$) and slightly improved by NG for HOD parameters ($5.6 \times 10^7 \rightarrow 3.9 \times 10^7$). This means the bulk of the improvement for the whole cosmo+HOD matrix comes from the change in the off-diagonal block, meaning the degeneracies between cosmological and HOD parameters. Visually inspecting the full Fisher matrices, I indeed found that several degeneracies are improved by NG, in particular, those between $w$ and HOD. This means that NG eases up the sensitivity of Dark Energy constraints on HOD parameters and possible modelling uncertainties. This comes from the structure of the covariance and cannot be mimicked, for example, by rescaling the Gaussian covariance by an arbitrary factor which would leave degeneracies untouched.

In conclusion, including braiding and in-survey covariances is a necessity for future high-density galaxy clustering analyses. This is both because it impacts error bars at a level above the precision requirements and also because it renders cosmological constraints more robust for astrophysical uncertainties.

\section*{Acknowledgements}
\vspace{0.2cm}

I thank Isaac Tutusaus for private communication on the  galaxy bias of the \Euclid{} expected photometric sample.
Part of this work was supported by funds of the D\'epartement de Physique Th\'eorique, Universit\'e de Gen\`eve. Part of this work was supported by a postdoctoral grant from Centre National d’Études Spatiales (CNES).

\bibliographystyle{aa}
\bibliography{bibliography}

\begin{thebibliography}{30}
\expandafter\ifx\csname natexlab\endcsname\relax\def\natexlab#1{#1}\fi

\bibitem[{{Abramo} {et~al.}(2015){Abramo}, {Balm{\`e}s}, {Lacasa}, \&
  {Lima}}]{Abramo2015}
{Abramo}, L.~R., {Balm{\`e}s}, I., {Lacasa}, F., \& {Lima}, M. 2015, \mnras,
  454, 2844

\bibitem[{{Akitsu} \& {Takada}(2017)}]{Akitsu2017}
{Akitsu}, K. \& {Takada}, M. 2017, ArXiv e-prints [\eprint[arXiv]{1711.00012}]

\bibitem[{{Barreira} {et~al.}(2018{\natexlab{a}}){Barreira}, {Krause}, \&
  {Schmidt}}]{Barreira2018b}
{Barreira}, A., {Krause}, E., \& {Schmidt}, F. 2018{\natexlab{a}}, \jcap, 10,
  053

\bibitem[{{Barreira} {et~al.}(2018{\natexlab{b}}){Barreira}, {Krause}, \&
  {Schmidt}}]{Barreira2018}
{Barreira}, A., {Krause}, E., \& {Schmidt}, F. 2018{\natexlab{b}}, \jcap, 6,
  015

\bibitem[{{Euclid Collaboration} {et~al.}(2019){Euclid Collaboration},
  {Blanchard}, {Camera}, {Carbone}, {Cardone}, {Casas}, {Ili{\'c}},
  {Kilbinger}, {Kitching}, {Kunz}, {Lacasa}, {Linder}, {Majerotto},
  {Markovi{\v{c}}}, {Martinelli}, {Pettorino}, {Pourtsidou}, {Sakr},
  {S{\'a}nchez}, {Sapone}, {Tutusaus}, {Yahia-Cherif}, {Yankelevich}, \&
  et~al.}]{Euclid-IST}
{Euclid Collaboration}, {Blanchard}, A., {Camera}, S., {et~al.} 2019, arXiv
  e-prints, arXiv:1910.09273

\bibitem[{{Harnois-D{\'e}raps} \& {Pen}(2013)}]{Harnois-Deraps2013}
{Harnois-D{\'e}raps}, J. \& {Pen}, U.-L. 2013, \mnras, 431, 3349

\bibitem[{{Hildebrandt} {et~al.}(2017){Hildebrandt}, {Viola}, {Heymans},
  {Joudaki}, {Kuijken}, {Blake}, {Erben}, {Joachimi}, {Klaes}, {Miller},
  {Morrison}, {Nakajima}, {Verdoes Kleijn}, {Amon}, {Choi}, {Covone}, {de
  Jong}, {Dvornik}, {Fenech Conti}, {Grado}, {Harnois-D{\'e}raps}, {Herbonnet},
  {Hoekstra}, {K{\"o}hlinger}, {McFarland}, {Mead}, {Merten}, {Napolitano},
  {Peacock}, {Radovich}, {Schneider}, {Simon}, {Valentijn}, {van den Busch},
  {van Uitert}, \& {Van Waerbeke}}]{Hildebrandt2017}
{Hildebrandt}, H., {Viola}, M., {Heymans}, C., {et~al.} 2017, \mnras, 465, 1454

\bibitem[{{Krause} \& {Eifler}(2017)}]{Krause2017}
{Krause}, E. \& {Eifler}, T. 2017, \mnras, 470, 2100

\bibitem[{{Krause} {et~al.}(2017){Krause}, {Eifler}, {Zuntz}, {Friedrich},
  {Troxel}, {Dodelson}, {Blazek}, {Secco}, \& {et al.}}]{Krause2017b}
{Krause}, E., {Eifler}, T.~F., {Zuntz}, J., {et~al.} 2017, ArXiv e-prints
  [\eprint[arXiv]{1706.09359}]

\bibitem[{{Lacasa}(2018)}]{Lacasa2018b}
{Lacasa}, F. 2018, \aap, 615, A1

\bibitem[{{Lacasa} \& {Grain}(2019)}]{Lacasa2019}
{Lacasa}, F. \& {Grain}, J. 2019, \aap, 624, A61

\bibitem[{{Lacasa} \& {Kunz}(2017)}]{Lacasa2017}
{Lacasa}, F. \& {Kunz}, M. 2017, \aap, 604, A104

\bibitem[{{Lacasa} {et~al.}(2018){Lacasa}, {Lima}, \& {Aguena}}]{Lacasa2018}
{Lacasa}, F., {Lima}, M., \& {Aguena}, M. 2018, \aap, 611, A83

\bibitem[{{Lacasa} {et~al.}(2014){Lacasa}, {P{\'e}nin}, \&
  {Aghanim}}]{Lacasa2014}
{Lacasa}, F., {P{\'e}nin}, A., \& {Aghanim}, N. 2014, \mnras, 439, 123

\bibitem[{{Lacasa} \& {Rosenfeld}(2016)}]{Lacasa2016}
{Lacasa}, F. \& {Rosenfeld}, R. 2016, \jcap, 8, 005

\bibitem[{Laureijs {et~al.}(2011)}]{Euclid-redbook}
Laureijs, R. {et~al.} 2011, ArXiv e-prints [\eprint[arXiv]{1110.3193}]

\bibitem[{{Li} {et~al.}(2014{\natexlab{a}}){Li}, {Hu}, \& {Takada}}]{Li2014}
{Li}, Y., {Hu}, W., \& {Takada}, M. 2014{\natexlab{a}}, \prd, 89, 083519

\bibitem[{{Li} {et~al.}(2014{\natexlab{b}}){Li}, {Hu}, \& {Takada}}]{Li2014b}
{Li}, Y., {Hu}, W., \& {Takada}, M. 2014{\natexlab{b}}, \prd, 90, 103530

\bibitem[{{Li} {et~al.}(2018){Li}, {Schmittfull}, \& {Seljak}}]{Li2018}
{Li}, Y., {Schmittfull}, M., \& {Seljak}, U. 2018, \jcap, 2, 022

\bibitem[{{Norberg} {et~al.}(2009){Norberg}, {Baugh}, {Gazta{\~n}aga}, \&
  {Croton}}]{Norberg2009}
{Norberg}, P., {Baugh}, C.~M., {Gazta{\~n}aga}, E., \& {Croton}, D.~J. 2009,
  \mnras, 396, 19

\bibitem[{{Planck Collaboration} {et~al.}(2018){Planck Collaboration},
  {Aghanim}, {Akrami}, {Ashdown}, {Aumont}, {Baccigalupi}, {Ballardini},
  {Banday}, {Barreiro}, {Bartolo}, \& et~al.}]{Planck2018-params}
{Planck Collaboration}, {Aghanim}, N., {Akrami}, Y., {et~al.} 2018, ArXiv
  e-prints [\eprint[arXiv]{1807.06209}]

\bibitem[{{Rizzato} {et~al.}(2019){Rizzato}, {Benabed}, {Bernardeau}, \&
  {Lacasa}}]{Rizzato2019}
{Rizzato}, M., {Benabed}, K., {Bernardeau}, F., \& {Lacasa}, F. 2019, \mnras,
  490, 4688

\bibitem[{{Sato} \& {Nishimichi}(2013)}]{Sato2013}
{Sato}, M. \& {Nishimichi}, T. 2013, \prd, 87, 123538

\bibitem[{{Sellentin} \& {Starck}(2019)}]{Sellentin2019}
{Sellentin}, E. \& {Starck}, J.-L. 2019, \jcap, 2019, 021

\bibitem[{{Takada} \& {Hu}(2013)}]{Takada2013}
{Takada}, M. \& {Hu}, W. 2013, \prd, 87, 123504

\bibitem[{{Tinker} {et~al.}(2008){Tinker}, {Kravtsov}, {Klypin}, {Abazajian},
  {Warren}, {Yepes}, {Gottl{\"o}ber}, \& {Holz}}]{Tinker2008}
{Tinker}, J., {Kravtsov}, A.~V., {Klypin}, A., {et~al.} 2008, \apj, 688, 709

\bibitem[{{Tinker} {et~al.}(2010){Tinker}, {Robertson}, {Kravtsov}, {Klypin},
  {Warren}, {Yepes}, \& {Gottl{\"o}ber}}]{Tinker2010}
{Tinker}, J.~L., {Robertson}, B.~E., {Kravtsov}, A.~V., {et~al.} 2010, \apj,
  724, 878

\bibitem[{{Troxel} {et~al.}(2018){Troxel}, {Krause}, {Chang}, {Eifler},
  {Friedrich}, {Gruen}, {MacCrann}, {Chen}, {Davis}, {DeRose}, {Dodelson},
  {Gatti}, {Hoyle}, {Huterer}, {Jarvis}, {Lacasa}, {Lemos}, {Peiris}, {Prat},
  {Samuroff}, {S{\'a}nchez}, {Sheldon}, {Vielzeuf}, {Wang}, {Zuntz}, {Lahav},
  {Abdalla}, {Allam}, {Annis}, {Avila}, {Bertin}, {Brooks}, {Burke}, {Carnero
  Rosell}, {Carrasco Kind}, {Carretero}, {Crocce}, {Cunha}, {D'Andrea}, {da
  Costa}, {De Vicente}, {Diehl}, {Doel}, {Evrard}, {Flaugher}, {Fosalba},
  {Frieman}, {Garc{\'\i}a-Bellido}, {Gaztanaga}, {Gerdes}, {Gruendl},
  {Gschwend}, {Gutierrez}, {Hartley}, {Hollowood}, {Honscheid}, {James},
  {Kirk}, {Kuehn}, {Kuropatkin}, {Li}, {Lima}, {March}, {Menanteau}, {Miquel},
  {Mohr}, {Ogando}, {Plazas}, {Roodman}, {Sanchez}, {Scarpine}, {Schindler},
  {Sevilla-Noarbe}, {Smith}, {Soares-Santos}, {Sobreira}, {Suchyta}, {Swanson},
  {Thomas}, {Walker}, \& {Wechsler}}]{Troxel2018}
{Troxel}, M.~A., {Krause}, E., {Chang}, C., {et~al.} 2018, \mnras, 479, 4998

\bibitem[{{Wadekar} \& {Scoccimarro}(2019)}]{Wadekar2019}
{Wadekar}, D. \& {Scoccimarro}, R. 2019, arXiv e-prints, arXiv:1910.02914

\bibitem[{{Zehavi} {et~al.}(2011){Zehavi}, {Zheng}, {Weinberg}, {Blanton},
  {Bahcall}, {Berlind}, {Brinkmann}, {Frieman}, {Gunn}, {Lupton}, {Nichol},
  {Percival}, {Schneider}, {Skibba}, {Strauss}, {Tegmark}, \&
  {York}}]{Zehavi2011}
{Zehavi}, I., {Zheng}, Z., {Weinberg}, D.~H., {et~al.} 2011, \apj, 736, 59

\end{thebibliography}

\appendix

\section{Redshift dependent Halo Occupation Distribution}\label{App:HOD}

The specification Eq.~\ref{Eq:Ngal(z)-Euclid} for the galaxy redshift distribution $n(z)$ corresponds to a magnitude-limited sample, and not a volume-limited sample as is normally required for a HOD analysis. To overcome this, I fit the HOD parameters at each redshift. As $n(z)$ is the only `data' available, I can fit a single HOD parameter and I\ have to assume relations for the others. Specifically, I fit the $M_\mr{min}$ parameter, assume that the ratio $M_\mr{ratio} = M_\mr{sat}/M_\mr{min}=10$ is constant and that $\sigma_{\log M}=0.5$ and $\alpha_\mr{sat}=1$ are constant. I then find that I can reproduce the fitted $M_\mr{min}(z)$ with a fourth order polynomial:
\ba
M_\mr{min}(z) = M_\mr{min}^a + M_\mr{min}^b \, z + M_\mr{min}^c \, z^2 + M_\mr{min}^d \, z^3 ,
\ea
with parameter values (rounded to the third decimal) $M_\mr{min}^a = 11.020$, $M_\mr{min}^b = -0.143$, $M_\mr{min}^c = 0.549$ and $M_\mr{min}^d = -0.105$.

\begin{figure}[!ht]
\begin{center}
\includegraphics[width=\linewidth]{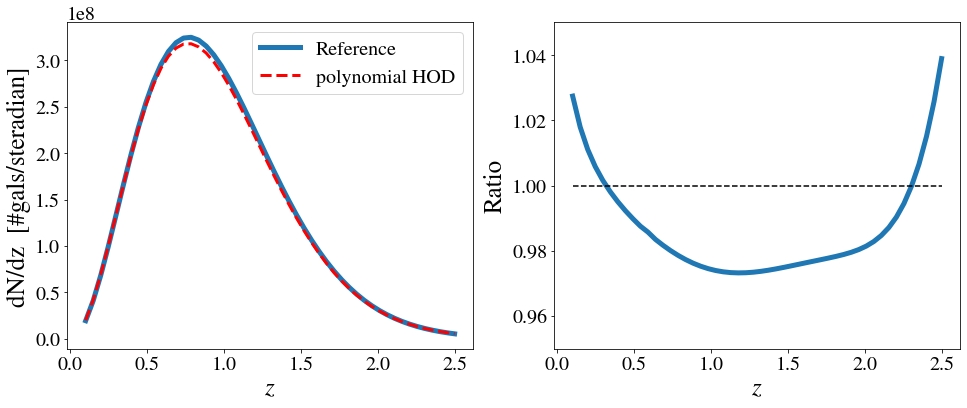}
\caption{\textit{Left:} specification for the galaxy redshift distribution from \cite{Euclid-redbook} (blue solid) and prediction from the polynomial HOD (red dashed). \textit{Right:} ratio of the two distributions.}
\label{Fig:Ngal_ref-vs-HOD}
\end{center}
\end{figure}

Using this polynomial redshift-dependent HOD (hereafter, polynomial HOD), Fig.~\ref{Fig:Ngal_ref-vs-HOD} shows the predicted galaxy redshift distribution $n(z)$ compared to the original specification.

\begin{figure}[!ht]
\begin{center}
\includegraphics[width=.9\linewidth]{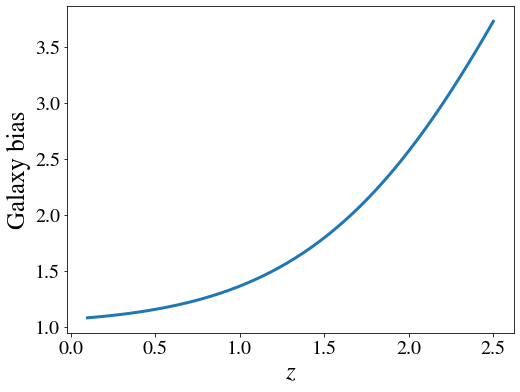}
\caption{Galaxy bias predicted from the polynomial HOD.}
\label{Fig:bgal-vs-z}
\end{center}
\end{figure}

We see that the redshift distribution is reproduced to better than 2.5\% accuracy over the whole redshift range. Now this means that given the galaxy bias at any order, the non-linear power spectrum (etc.) can all be predicted without any additional free parameter. For instance, Fig.~\ref{Fig:bgal-vs-z} shows the predicted first order galaxy bias.

I find  agreement between this galaxy bias and preliminary results from \Euclid{} internal simulations (Isaac Tutusaus, private communication), which shows that this simple parametrisation is indeed capable of capturing the redshift evolution of the expected \Euclid{} photometric sample to a satisfactory extent.

\end{document}